\documentclass[12pt]{article}
\usepackage{epsf}
\newcommand{\sect}[1]{\section{#1}\setcounter{equation}{0}}

\newcommand{\II}{\leavevmode\hbox{\rm{\small1\kern-3.8pt\normalsize1}}}

\def\Box{{\hbox{$\sqcup$}\llap{\hbox{$\sqcap$}}}}

\begin{document}
\addtolength{\baselineskip}{1.5mm}
\renewcommand{\thefootnote}{\fnsymbol{footnote}}

\begin{flushright}
DAMTP--1998--160\\
gr-qc/9812032\\
\end{flushright}
\smallskip

\begin{center}
{\huge{Bounds on negative energy densities in static \\[2mm]
        space-times}}\\[12mm]
{Christopher J. Fewster\footnote{Electronic address: {\tt cjf3@york.ac.uk}}}\\[6mm]
{\sl Department of Mathematics, University of York, Heslington,
York YO10 5DD, England}\\[8mm]

{Edward Teo\footnote{Electronic address: {\tt E.Teo@damtp.cam.ac.uk}}}\\[6mm]
{\sl Department of Applied Mathematics and Theoretical Physics,
University of Cambridge,\\
Silver Street,
Cambridge CB3 9EW,
England\\[3mm] 
Department of Physics, National University of Singapore, 
Singapore 119260}
\end{center}
\vspace{0.8cm}

\centerline{\bf Abstract}\bigskip
\noindent
Certain exotic phenomena in general relativity, such as backward time 
travel, appear to require the presence of matter with negative energy. 
While quantum fields are a possible source of negative energy densities, 
there are lower bounds---known as quantum inequalities---that constrain 
their duration and magnitude. In this paper, we derive new quantum 
inequalities for scalar fields in static space-times, as measured by 
static observers with a choice of sampling function. Unlike those 
previously derived by Pfenning and Ford, our results do not assume any 
specific sampling function. We then calculate these bounds in static 
three- and four-dimensional Robertson--Walker universes, the de Sitter
universe, and the Schwarzschild black hole. In each case, the new 
inequality is stronger than that of Pfenning and Ford for their 
particular choice of sampling function. 

\vfill\eject


\setcounter{footnote}{0}
\renewcommand{\thefootnote}{\arabic{footnote}}

\sect{Introduction}

In recent years, there has been much interest in various exotic solutions
of general relativity---such as traversable wormholes \cite{MT,MTY}, the 
Alcubierre ``warp drive'' \cite{alcubierre}, and the Krasnikov ``tube''
\cite{kransnikov}---that permit hyperfast or backward time travel. However, 
these space-times without exception require the presence of matter which 
possess {\it negative\/} energy densities \cite{FR-worm,PF-warp,ER,olum}, 
and hence violate the standard energy conditions. 

Now, it is well-known that quantum field theory, unlike classical physics, 
allows energy density to be unboundedly negative at a point in space-time
\cite{epstein}. Should the theory place no restrictions on this negative 
energy, quantum fields could be used to produce gross macroscopic effects 
such as those mentioned above, or even a violation of cosmic censorship 
or the second law of thermodynamics. It is therefore important 
to have a quantitative handle on the permitted amount of negative energy 
in a neighbourhood of a space-time point. 

Ford and Roman \cite{FRa,FR} have found inequalities which constrain the 
duration and magnitude of negative energy densities for quantised free,
real scalar fields in Minkowski space. They show that a static observer, 
who samples the energy density by time-averaging it against the Lorentzian 
function
\begin{equation}
\label{Lorentzian}
f(t)={t_0\over\pi}{1\over t^2+t_0^2}\,,
\end{equation}
obtains a result which is bounded from below by a negative quantity
depending inversely on the characteristic timescale $t_0$. For example,
in the case of a massless scalar field in four dimensions, the
renormalised energy density in any quantum state satisfies
\begin{equation}
\rho\geq-{3\over32\pi^2t_0^4}\,.
\end{equation}
This means the more negative the energy density that is present in an
interval, the shorter the duration of this interval must be. Thus, this
``quantum inequality''---in a way reminiscent of the uncertainty principle
of quantum mechanics\footnote{However, the derivation of the
quantum inequalities does not depend on any putative time-energy
uncertainty principle.}---serves to limit any large-scale, long-time
occurrence of negative energy. In the infinite sampling time limit
$t_0\rightarrow\infty$, it reduces to the usual averaged weak energy 
condition (for quantum fields \cite{PFb,pfenning}).

Eveson and one of the present authors~\cite{fewster} have recently presented
a different derivation of the quantum inequalities for a massive scalar
field in $n$-dimensional Minkowski space (with $n\geq2$). The method used is
straightforward---involving only the canonical commutation relations and
the convolution theorem of Fourier analysis---and has the virtue of being valid 
for any smooth, non-negative and even sampling function decaying sufficiently 
quickly at infinity. Furthermore, the resulting bounds turn out to
be stronger than those obtained by Ford and Roman~\cite{FRa,FR} when the
Lorentzian sampling function is applied.

In the present paper, we extend this method to derive quantum inequalities 
for scalar fields in generally curved but static space-times using
arbitrary smooth, non-negative (although not necessarily even, as assumed 
in \cite{fewster}) sampling functions of sufficiently rapid decay.
We obtain a lower bound on the averaged normal-ordered energy density in
the Fock space built on the static vacuum in terms of the appropriate
mode functions. Since the normal-ordered energy density in a given state
is the difference between the renormalised energy density in this state
and the (generally nonzero and potentially negative) renormalised energy
density of the static vacuum, our bound also constrains the renormalised
energy density (cf.~\cite{PFb}).

We apply our bound to
several examples where the bound can be explicitly evaluated, namely the
three- and four-dimensional Robertson--Walker universes, the de Sitter
universe, and the Schwarzschild black hole. In all these cases, we obtain 
bounds which are up to an order of magnitude stronger than those previously 
derived by Pfenning and Ford \cite{PFb,pfenning,PFa} for the specific 
sampling function they used. 

\sect{Derivation of the quantum inequality}
\label{QI}

We shall consider $n+1$-dimensional space-times that are globally static,
with time-like Killing vector $\partial_t$. The metric of such a space-time
takes the general form
\begin{equation}
{\rm d}s^2=-|g_{tt}({\bf x})|{\rm d}t^2+
g_{ij}({\bf x}){\rm d}x^i{\rm d}x^j,
\end{equation}
where ${\bf x}=(x^1,x^2\dots,x^n)$ and $i,j=1,2,\dots,n$. The equation of a 
free, real scalar field $\phi$ of mass $\mu\geq0$ in this space-time is 
\begin{equation}
-{1\over|g_{tt}|}\partial_t^2\phi+\nabla^i\nabla_i\phi
-\mu^2\phi=0\,.
\end{equation}
Suppose it admits a complete, orthonormal set of positive frequency 
solutions. We write these mode functions as
\begin{equation}
f_\lambda(t,{\bf x})=U_\lambda({\bf x}){\rm e}^{-i\omega_\lambda t},
\end{equation}
where $\lambda$ denotes the set of quantum numbers needed to specify the 
mode (which may be continuous or discrete). A general quantum scalar field 
can then be expanded as
\begin{equation}
\phi=\sum_\lambda(a_\lambda f_\lambda+a_\lambda^\dagger
f_\lambda^\ast)\,,
\end{equation}
in terms of creation and annihilation operators $a_\lambda^\dagger$, 
$a_\lambda$ obeying the canonical commutation relations
\begin{equation}
\label{ccr}
[a_\lambda,a_{\lambda^\prime}^\dagger]=\delta_{\lambda
\lambda^\prime}\II\,,\qquad
[a_\lambda,a_{\lambda^\prime}]=[a_\lambda^\dagger,
a_{\lambda^\prime}^\dagger]=0\,,
\end{equation}
and which generate the Fock space built on the static vacuum state
$|0\rangle$.
We shall be interested in the energy density of $\phi$ 
along the world-line $x^\mu(t)=(t,{\bf x_0})$ of a {\em static\/} observer,
with ${\bf x_0}$ kept fixed. If the field is in a normalised quantum state 
$|\psi\rangle$, the normal-ordered energy density as measured 
by such an observer at time $t$ is \cite{PFb,pfenning}
\begin{eqnarray}
\langle\,:T_{\mu\nu}u^\mu u^\nu:\,\rangle&=&{\rm Re}\,
\sum_{\lambda,\lambda^\prime}\bigg\{{\omega_\lambda\omega_{\lambda^\prime}
\over|g_{tt}|}\Big[U_\lambda^\ast U_{\lambda^\prime}\langle a_\lambda^\dagger 
a_{\lambda^\prime}\rangle{\rm e}^{i(\omega_\lambda-\omega_{\lambda^\prime})t}
-U_\lambda U_{\lambda^\prime}\langle a_\lambda a_{\lambda^\prime}\rangle
{\rm e}^{-i(\omega_\lambda+\omega_{\lambda^\prime})t}\Big]\cr
&&\hskip.5in+\Big[\nabla^iU_\lambda^\ast \nabla_i
U_{\lambda^\prime}\langle a_\lambda^\dagger a_{\lambda^\prime}\rangle
{\rm e}^{i(\omega_\lambda-\omega_{\lambda^\prime})t}
+\nabla^iU_\lambda\nabla_iU_{\lambda^\prime}\langle a_\lambda 
a_{\lambda^\prime}\rangle
{\rm e}^{-i(\omega_\lambda+\omega_{\lambda^\prime})t}\Big]\cr
&&\hskip.5in+m^2\Big[U_\lambda^\ast 
U_{\lambda^\prime}\langle a_\lambda^\dagger a_{\lambda^\prime}\rangle{\rm 
e}^{i(\omega_\lambda-\omega_{\lambda^\prime})t}+U_\lambda U_{\lambda^\prime} 
\langle a_\lambda a_{\lambda^\prime}\rangle
{\rm e}^{-i(\omega_\lambda+\omega_{\lambda^\prime})t}\Big]\bigg\}\,,
\label{Tmunuexp}
\end{eqnarray}
where $u^\mu=\big(|g_{tt}|^{-1/2},{\bf 0}\big)$ is 
the observer's four-velocity, and $U_\lambda$ and its derivatives are
evaluated at ${\bf x_0}$. We have also written
$\langle\,\cdot\,\rangle\equiv\langle\psi|\cdot|\psi\rangle$
for brevity. Recall that the normal-ordered energy density is the difference between
the renormalised energy density in the two states $|\psi\rangle$ and
$|0\rangle$.

We now define a weighted energy density
\begin{equation}
\rho = \int_{-\infty}^\infty{\rm d}t\,\langle\,:T_{\mu\nu}u^\mu u^\nu
:\,\rangle\, f(t)\,,
\end{equation}
where $f$ is any smooth, non-negative function decaying at least as fast as
${\rm O}(t^{-2})$ at infinity, and normalised to have unit integral. 
Ford and coworkers~\cite{FRa,FR,PFb,pfenning,PFa} employ
the Lorentzian function~(\ref{Lorentzian}), whose specific properties
play a key r\^ole in their arguments [in particular, the Fourier transform
of~(\ref{Lorentzian}) is simply the function $\exp(-|\omega|t_0)$]; we
emphasise that our arguments apply to general $f$. Substituting from
Eq.~(\ref{Tmunuexp}), the weighted energy density 
measured by the observer is 
\begin{eqnarray}
\rho &=&{\rm Re}\,
\sum_{\lambda,\lambda^\prime}\bigg\{{\omega_\lambda\omega_{\lambda^\prime}
\over|g_{tt}|}\Big[U_\lambda^\ast U_{\lambda^\prime}\langle a_\lambda^\dagger 
a_{\lambda^\prime}\rangle\widehat{f}(\omega_{\lambda'}-\omega_\lambda)
-U_\lambda U_{\lambda^\prime}\langle a_\lambda a_{\lambda^\prime}\rangle
\widehat{f}(\omega_\lambda+\omega_{\lambda'})\Big]\cr
&&\hskip.5in+\Big[\nabla^iU_\lambda^\ast \nabla_i
U_{\lambda^\prime}\langle a_\lambda^\dagger a_{\lambda^\prime}\rangle
\widehat{f}(\omega_{\lambda'}-\omega_\lambda)
+\nabla^iU_\lambda\nabla_iU_{\lambda^\prime}\langle a_\lambda 
a_{\lambda^\prime}\rangle
\widehat{f}(\omega_\lambda+\omega_{\lambda'})
\Big]\cr
&&\hskip.5in+m^2\Big[U_\lambda^\ast 
U_{\lambda^\prime}\langle a_\lambda^\dagger a_{\lambda^\prime}\rangle
\widehat{f}(\omega_{\lambda'}-\omega_\lambda)
+U_\lambda U_{\lambda^\prime} 
\langle a_\lambda a_{\lambda^\prime}\rangle
\widehat{f}(\omega_\lambda+\omega_{\lambda'})\Big]\bigg\}\,,
\end{eqnarray}
where we define the Fourier transform of $f$ by
\begin{equation}
\widehat{f}(\omega)=\int_{-\infty}^\infty{\rm d}t\,f(t){\rm 
e}^{-i\omega t}\,.
\end{equation}

By applying the inequality (\ref{inequality}), proved in the Appendix, 
to each of the cases
$q_\lambda={\omega_\lambda\over|g_{tt}|^{1/2}}U_\lambda$, $\nabla_iU_\lambda$
and $mU_\lambda$, we obtain the following manifestly negative lower bound for $\rho$:
\begin{equation}
\rho\geq-{1\over2\pi}\int_0^\infty{\rm d}\omega\,\sum_\lambda
\bigg({\omega_\lambda^2\over|g_{tt}|}U_\lambda^\ast U_\lambda
+\nabla^iU_\lambda^\ast\nabla_iU_\lambda+m^2U_\lambda^\ast U_\lambda\bigg)
\left|\widehat{f^{1/2}}(\omega+\omega_\lambda)\right|^2.
\end{equation}
Using the field equation satisfied by the spatial mode function 
\cite{PFb,pfenning}:
\begin{equation}
\nabla^i\nabla_iU_\lambda+\bigg({\omega_\lambda^2\over|g_{tt}|}
-m^2\bigg)U_\lambda=0\,,
\end{equation}
this inequality can be rewritten as
\begin{equation}
\label{QIb}
\rho\geq-{1\over\pi}\int_0^\infty{\rm d}\omega\,\sum_\lambda
\bigg({\omega_\lambda^2\over|g_{tt}|}+{1\over4}\nabla^i\nabla_i\bigg)
|U_\lambda|^2\left|\widehat{f^{1/2}}(\omega+\omega_\lambda)\right|^2.
\end{equation}
This is the desired quantum inequality, which is valid for general
sampling functions $f(t)$, subject to the above-stated conditions. 
Another useful form of it can be obtained by 
introducing the new variable $u=\omega+\omega_\lambda$:
\begin{equation}
\label{QIc}
\rho\geq-{1\over\pi}\int_{\omega_{\rm min}}^\infty{\rm d}u\,
\left|\widehat{f^{1/2}}(u)\right|^2\sum_{\lambda~{\rm s.t.}~\omega_\lambda\leq u}
\bigg({\omega_\lambda^2\over|g_{tt}|}+{1\over4}\nabla^i\nabla_i\bigg)
\,|U_\lambda|^2,
\end{equation}
with $\omega_{\rm min}\equiv\min_\lambda\omega_\lambda$.

To simplify it any further would require a specific choice of $f(t)$. For
example, with the even sampling function 
\begin{equation}
\label{mysamp}
f(t)={2\over\pi}{t_0^3\over(t^2+t_0^2)^2}\,,
\end{equation}
that is peaked at $t=0$, we have 
\begin{equation}
\left|\widehat{f^{1/2}}(\omega)\right|^2=2\pi t_0{\rm e}^{-2|\omega|t_0}.
\end{equation}
In this case, the quantum inequality can be expressed in terms of the 
Euclidean Green's function
\begin{equation}
G_{\rm E}(t,{\bf x};t^\prime,{\bf x}^\prime)=\sum_\lambda
U_\lambda^\ast({\bf x})U_\lambda({\bf x}^\prime){\rm e}^{\omega_\lambda
(t-t^\prime)},
\end{equation}
quite compactly as
\begin{equation}
\rho\geq-\hbox{$1\over4$}\Box_{\rm E}G_{\rm E}(-t_0,{\bf x};
t_0,{\bf x})\,,
\end{equation}
where $\Box_{\rm E}\equiv{1\over|g_{tt}|}\partial_{t_0}^2+\nabla^i\nabla_i$ 
is the Euclidean wave operator. This bound is, in
fact, identical to one that was derived in \cite{PFb,pfenning} assuming the
Lorentzian sampling function (\ref{Lorentzian}). But because (\ref{mysamp})
is a more sharply peaked function [half the area under the Lorentzian function
lies within $|t|<t_0$, while this figure is
${1\over2}+{1\over\pi}\simeq0.82$ for
(\ref{mysamp})], this is a first indication that the
inequality derived here is a stronger result.

Finally, we record the fact that for the Lorentzian function,
\begin{equation}
\label{FT}
\left|\widehat{f^{1/2}}(\omega)\right|^2=\frac{4t_0}{\pi}
K_0(t_0|\omega|)^2,
\end{equation}
where $K_0(x)$ is the modified Bessel function of zeroth order.
In the rest of this paper, we shall consider the quantum inequality 
in specific examples of globally static space-times where the left-hand
side of (\ref{QIb}) or (\ref{QIc}) can be explicitly evaluated.
As these examples have been considered previously by Pfenning and Ford 
\cite{PFb,pfenning,PFa}, we shall at times be brief and refer to their
papers for more details. For the most part we will closely follow their
notation and conventions.

\sect{Minkowski space}

We begin with a review of the quantum inequality in $n+1$-dimensional
Minkowski space, the case that was treated in \cite{fewster}. The mode 
functions for a free scalar field of mass $\mu$ are 
\begin{equation}
U_{\bf k}({\bf x})={1\over[(2\pi)^n
2\omega_{\bf k}]^{1/2}}{\rm e}^{i{\bf k}\cdot{\bf x}},\qquad
\omega_{\bf k}=\sqrt{|{\bf k}|^2+\mu^2}\,,
\end{equation}
with each component of the $n$-dimensional (spatial) momentum covector ${\bf k}$
satisfying $-\infty<k_i<\infty$. The quantum inequality (\ref{QIb}) becomes
\begin{eqnarray}
\label{Mbound}
\rho&\geq&-{1\over2\pi}\int_0^\infty{\rm d}\omega
\int{{\rm d}^n{\bf k}\over(2\pi)^n}\,\omega_{\bf k}
\left|\widehat{f^{1/2}}(\omega+\omega_{\bf k})\right|^2\cr
&=&-\frac{C_n}{2\pi}\int_0^\infty{\rm d}\omega\int_\mu^\infty{\rm d}\omega'\,
{\omega'}^2({\omega'}^2-\mu^2)^{n/2-1}
\left|\widehat{f^{1/2}}(\omega+\omega')\right|^2,
\end{eqnarray}
where $C_n$ is equal to the area of the unit $n-1$-sphere divided by
$(2\pi)^n$, i.e.,
\begin{equation}
C_n\equiv\frac{1}{2^{n-1}\pi^{n/2}\Gamma(\frac{1}{2}n)}\,.
\end{equation}
If we make the change of variables $u=\omega+\omega'$ and $v=\omega'$, 
the quantum inequality (\ref{Mbound}) can be rewritten as
\begin{equation}
\label{Mbound2}
\rho\ge-\frac{C_n}{2\pi(n+1)}\int_\mu^\infty{\rm d}u\,
\left|\widehat{f^{1/2}}(u)\right|^2 u^{n+1}Q_n\bigg(\frac{u}{\mu}\bigg),
\end{equation}
where the functions $Q_n(x)$ are defined by
\begin{equation}
Q_n(x) = (n+1)x^{-(n+1)}\int_1^x{\rm d}y\, y^2(y^2-1)^{n/2-1}.
\end{equation}

There are several special cases in which this bound can be evaluated 
analytically \cite{fewster}, notably massless fields in two and 
four dimensions with the sampling function 
(\ref{Lorentzian}). In the former, the bound is four times stronger 
than that derived by Ford and Roman \cite{FR}, but $1{1\over2}$ times weaker 
than the optimal one of Flanagan \cite{flanagan}. In the latter case, the 
present bound is $9\over64$ of Ford and Roman's result.

\sect{Three-dimensional closed universe}
\label{IIIDCU}

The line element for the static, three-dimensional closed universe is
\begin{equation}
{\rm d}s^2=-{\rm d}t^2+a^2({\rm d}\theta^2+\sin^2\theta\,{\rm d}\varphi^2)\,,
\end{equation}
where $a$ is the radius of the two-sphere at each constant time-slice,
and the angular variables take values $0\leq\theta\leq\pi$,
$0\leq\varphi<2\pi$ (and will do so for all the space-times considered in
this paper). We 
consider the massive scalar field equation on this background with a 
coupling of strength $\xi$ to the scalar curvature $R=2/a^2$:
\begin{equation}
\Box\phi-(\mu^2+\xi R)\phi=0\,,
\end{equation}
whose mode-function solutions are given in terms of the usual spherical
harmonics $Y_{lm}(\theta,\varphi)$ by~\cite{pfenning,PFa}
\begin{equation}
U_{lm}({\bf x})={1\over(2\omega_la^2)^{1/2}}Y_{lm}(\theta,\varphi)\,,
\end{equation}
for $l=0,1,2,\dots$ and $m=-l,-l+1,\dots,l$, with
\begin{equation}
\omega_l=a^{-1}\sqrt{l(l+1)+2\xi+(a\mu)^2}\, .
\end{equation}
The $Y_{lm}(\theta,\varphi)$ obey the sum rule
\begin{equation}
\label{sumrule}
\sum_{m=-l}^{l}\big|Y_{lm}(\theta,\varphi)\big|^2={2l+1\over4\pi}\,,
\end{equation}
which can be used in (\ref{QIb}) and (\ref{QIc}) to show that
\begin{eqnarray}
\label{3Dbound}
\rho&\geq&-{1\over8\pi^2a^2}\int_0^\infty{\rm d}\omega\sum_{l=0}^\infty\,
(2l+1)\omega_l\left|\widehat{f^{1/2}}(\omega+\omega_l)\right|^2\cr
&=&-{1\over8\pi^2a^2}\int_{\omega_0}^\infty{\rm d}u\,
\left|\widehat{f^{1/2}}(u)\right|^2
\sum_{l=0}^{N(u)}\,(2l+1)\omega_l\,.
\end{eqnarray}
Here, $N(u)\equiv\max\{n\in{\bf Z}:\,\omega_n\leq u\}$, i.e.,
\begin{equation}
N(u)=\left\lfloor{\sqrt{1-4[2\xi+(a\mu)^2-(au)^2]}-1\over2}
\right\rfloor,
\end{equation}
where $\lfloor x \rfloor$ denotes the integer part of $x$.

While the bound in (\ref{3Dbound}) can be readily evaluated using 
numerical techniques, it may be worthwhile to first simplify it 
analytically as much as possible. This may be useful if one should
want to draw conclusions about its general properties. In particular,
we shall present a general strategy for approximating finite summations 
like that in (\ref{3Dbound}).

\begin{figure}[t]
\begin{center}
\epsfxsize=6.in
\epsffile{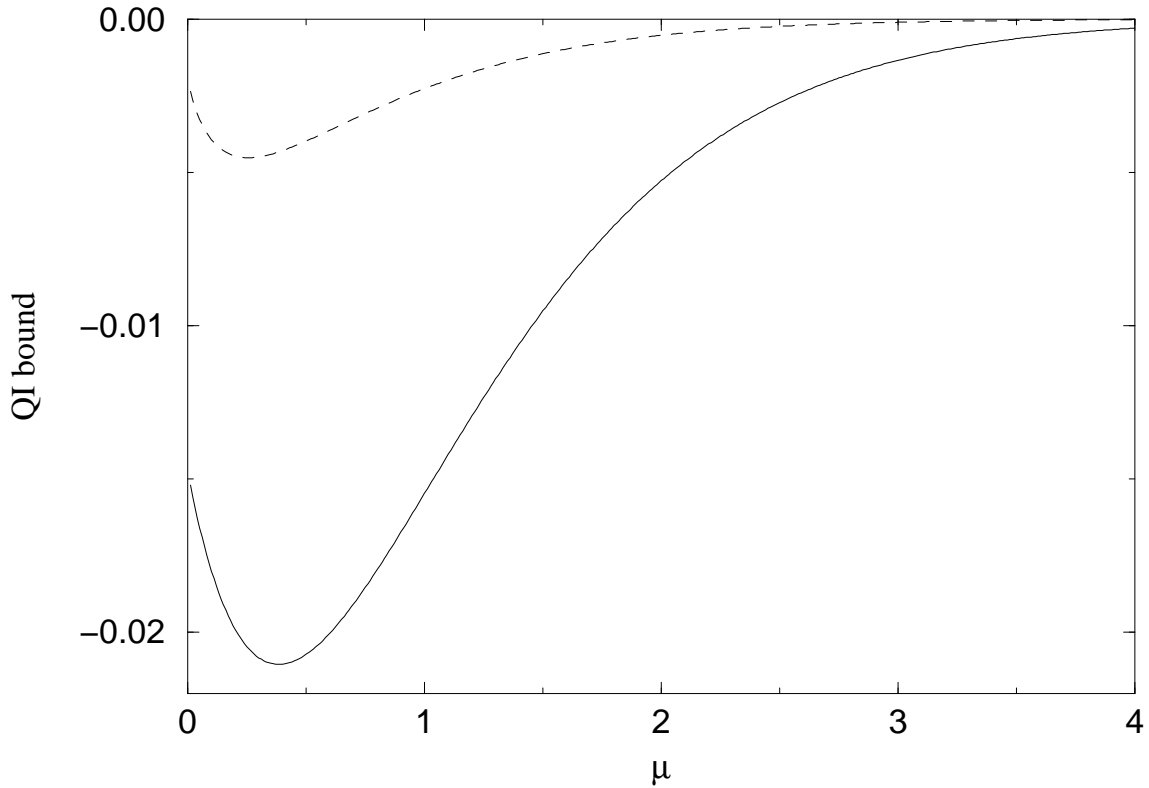}
\caption{Graph of the QI bound for the 3D
closed universe [dashed line], and that obtained by Pfenning and Ford 
[solid line], against $\mu$.}
\label{fig2}
\end{center}
\end{figure}

The summation in (\ref{3Dbound}) can be evaluated using the 
trapezoidal rule of numerical integration (e.g., see Eq.~(3.6.1)
of \cite{Hildebrand}):
\begin{equation}
\label{trapez}
\sum_{n=0}^Ng(n)=\int_0^N{\rm d}x\,g(x)+{1\over2}\left[g(0)+g(N)\right]
+{N\over12}g''(\zeta)\,,
\end{equation}
for some $\zeta\in(0,N)$. In the present case, $g(x)=(2x+1)\sqrt{x(x+1)+
2\xi+(a\mu)^2}$, and the integral in (\ref{trapez}) can be evaluated
analytically. Furthermore, $g''(\zeta)$ is non-decreasing in
the interval in question, so its occurrence in (\ref{3Dbound}) can be
replaced by $g''(N)$, at the expense of weakening the bound slightly. 
We obtain the final inequality
\begin{equation}
\label{3Dbound1}
\rho\geq-{1\over8\pi^2a^3}\int_{\omega_0}^\infty{\rm d}u\,
\left|\widehat{f^{1/2}}(u)\right|^2
\left\{\int_0^{N(u)}{\rm d}x\,g(x)+{1\over2}\left[g(0)+g(N(u))\right]
+{N(u)\over12}g''(N(u))\right\},
\end{equation}
with
\begin{eqnarray}
\int{\rm d}x\,g(x)&=&{2\over3}\left[x(x+1)+2\xi+(a\mu)^2\right]^{3/2},\cr
g''(x)&=&{3(2x+1)\over\sqrt{x(x+1)+2\xi+(a\mu)^2}}-{1\over4}
{(2x+1)^3\over\big[x(x+1)+2\xi+(a\mu)^2\big]^{3/2}}\,.
\end{eqnarray}

The graph of the bound in (\ref{3Dbound}) is plotted against mass
in Fig.~\ref{fig2}, for $a=1$ and $\xi=0$. As usual, the sampling 
function $f(t)$ is taken to be the Lorentzian function 
(\ref{Lorentzian}), with $t_0=1$. When plotted on the same scale, that 
of (\ref{3Dbound1}) is almost indistinguishable from the former graph. 
For comparison, the corresponding bound obtained by Pfenning and Ford  
\cite{pfenning,PFa} is also plotted in  Fig.~\ref{fig2}. It is clear that 
our bound is stronger for all values of mass.

\sect{Four-dimensional Robertson--Walker universe}
\label{IVDRWU}

We shall first consider the case of the open universe, before proceeding
to the closed universe. The line element is
\begin{equation}
{\rm d}s^2=-{\rm d}t^2+a^2\left[{\rm d}\chi^2+\sinh^2\chi\,
\big({\rm d}\theta^2+\sin^2\theta\,{\rm d}\varphi^2\big)\right],
\end{equation}
where $a$ characterises the scale of the spatial section, and 
$0\leq\chi<\infty$. 
The mode functions for a scalar field of mass $\mu$ are \cite{parker}
\begin{eqnarray}
U_{qlm}({\bf x})&=&{1\over(2\omega_qa^3)^{1/2}} 
\Pi^{(-)}_{ql}(\chi)Y_{lm}(\theta,\varphi)\,,\cr
\omega_q&=&\sqrt{{q^2+1\over a^2}+\mu^2}\,,
\end{eqnarray}
with $0<q<\infty$ and $l,m$ as usual. The functions $\Pi^-_{ql}(\chi)$ satisfy
\begin{equation}
\Pi^{(-)}_{ql}(\chi)\propto\sinh^l\chi\,
\bigg({{\rm d}\over{\rm d}\cosh\chi}\bigg)^{l+1}\cos q\chi\,,
\end{equation}
and obey the sum rule
\begin{equation}
\sum_{l,m}\,\left|\Pi^{(-)}_{ql}(\chi)Y_{lm}
(\theta,\varphi)\right|^2 ={q^2\over 2\pi^2}\,.
\end{equation}
The right-hand side does not depend on the angular variables, as is 
expected of a system with isotropic symmetry. Hence, the quantum inequality
(\ref{QIb}) becomes
\begin{equation}
\label{Cbound}
\rho\geq-{1\over4\pi^3a^3}\int_0^\infty{\rm d}\omega
\int_0^\infty{\rm d}q\,\omega_qq^2
\left|\widehat{f^{1/2}}(\omega+\omega_q)\right|^2.
\end{equation}

Note that this bound is identical in form to that in (four-dimensional)
Minkowski space. Both (\ref{Mbound}) and (\ref{Cbound}) can, in fact, be 
written as
\begin{eqnarray}
\rho&\geq&-{1\over4\pi^3}\int_0^\infty{\rm d}\omega
\int_C^\infty{\rm d}\omega'\,\omega'^2\sqrt{\omega'^2-C^2}
\left|\widehat{f^{1/2}}(\omega+\omega')\right|^2\cr
&=&-\frac{1}{16\pi^3}\int_C^\infty{\rm d}u\,\left|\widehat{f^{1/2}}(u)\right|^2 
u^4 Q_3\left(\frac{u}{C}\right),
\end{eqnarray}
where
\begin{equation}
C\equiv\sqrt{{\epsilon\over a^2}+\mu^2},\qquad\epsilon=
\cases{0&Minkowski space;\cr1&open universe,}
\end{equation}
and an explicit expression (and graph) for $Q_3(x)$ can be found in 
\cite{fewster}. The Minkowski space result is obviously recovered in 
the limit of infinite $a$. Furthermore, since $Q_3(x)$ is an increasing
function on $[1,\infty)$, the bound for general $a$ is tighter than that
in Minkowski space for all sampling functions $f(t)$. Pfenning and Ford 
\cite{pfenning,PFa} also noted this for their particular choice of $f(t)$. 

We now turn to the closed or Einstein universe, with line element
\begin{equation}
{\rm d}s^2=-{\rm d}t^2+a^2\left[{\rm d}\chi^2+\sin^2\chi\,
\big({\rm d}\theta^2+\sin^2\theta\,{\rm d}\varphi^2\big)\right],
\end{equation}
where $0\leq\chi\leq\pi$. The mode functions are \cite{parker}
\begin{eqnarray}
U_{nlm}({\bf x})&=&{1\over(2\omega_na^3)^{1/2}} 
\Pi^{(+)}_{nl}(\chi)\, Y_{lm}(\theta,\varphi)\,,\cr
\omega_n&=&\sqrt{{n(n+2)\over a^2}+\mu^2}\,,
\end{eqnarray}
with $n=0,1,2,\dots$, $l=0,1,\dots,n$, $m=-l,-l+1\ldots,l$, and 
\begin{equation}
\Pi^{(+)}_{nl}(\chi)\propto\sin^l\chi\,
\bigg({{\rm d}\over{\rm d}\cos\chi}\bigg)^{l+1}\cosh (n+1)\chi\,.
\end{equation}
Using the sum rule
\begin{equation}
\sum_{l,m}\,\left|\Pi^{(+)}_{nl}(\chi)Y_{lm}
(\theta,\varphi)\right|^2={(n+1)^2\over 2\pi^2}\,,
\end{equation}
we obtain the quantum inequality
\begin{eqnarray}
\label{RWbound}
\rho&\geq&-{1\over4\pi^3a^3}\int_0^\infty{\rm d}\omega
\sum_{n=0}^\infty\,\omega_n(n+1)^2
\left|\widehat{f^{1/2}}(\omega+\omega_n)\right|^2\cr
&=&-{1\over4\pi^3a^3}\int_\mu^\infty{\rm d}u\,
\left|\widehat{f^{1/2}}(u)\right|^2
\sum_{n=0}^{N(u)}\,\omega_n(n+1)^2,
\end{eqnarray}
with
\begin{equation}
N(u)\equiv\Big\lfloor\sqrt{(au)^2-(a\mu)^2+1}-1\Big\rfloor\,.
\end{equation}

\begin{figure}[t]
\begin{center}
\epsfxsize=6.in
\epsffile{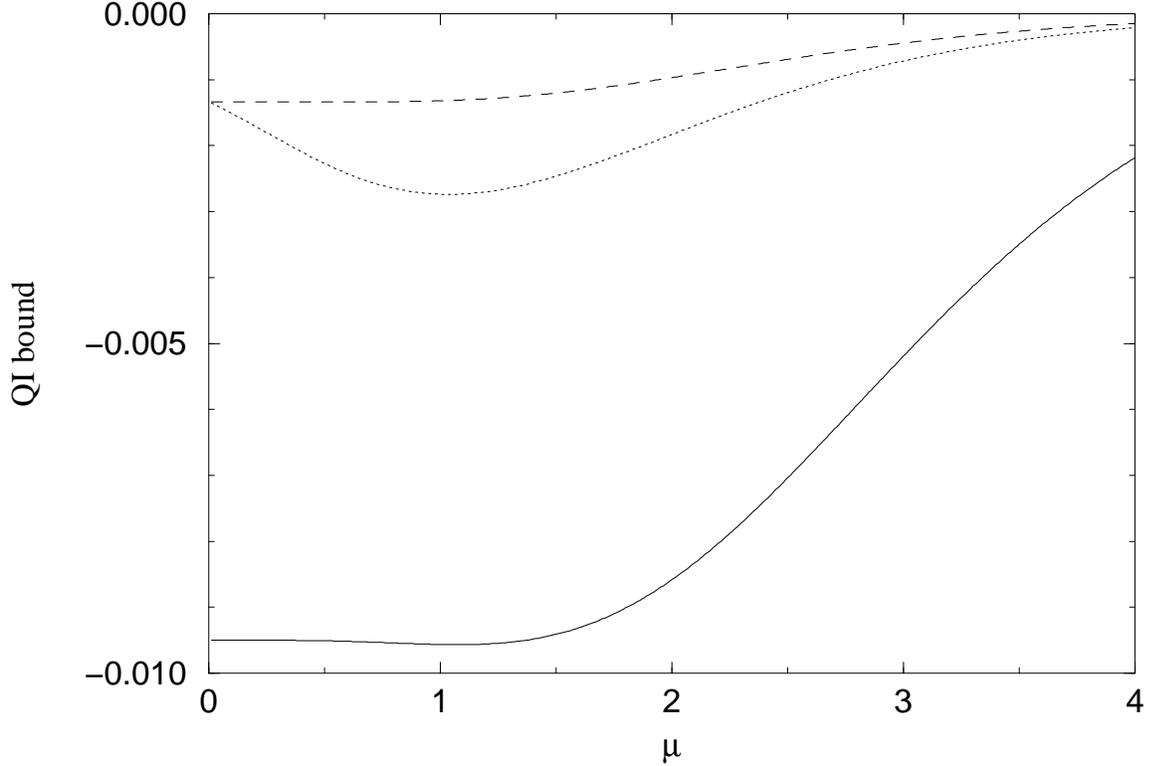}
\caption{Graphs of the QI bounds for the 4D
closed universe with $a\mu=1$: (\ref{Boundmua1}) using a dashed line, 
(\ref{Boundmua2}) using a dotted line, and that obtained by Pfenning and
Ford [solid line].}
\label{fig3}
\end{center}
\end{figure}

An obvious special case to investigate is $a\mu=1$, in which
$\omega_n=\mu(n+1)$. The sum in (\ref{RWbound}) may then be evaluated exactly, 
to give
\begin{equation} 
\label{Boundmua1}
\rho \ge -\frac{1}{16\pi^3a^4}\int_\mu^\infty{\rm d}u\,
\left|\widehat{f^{1/2}}(u)\right|^2(N(u)+1)^2(N(u)+2)^2.
\end{equation}
This bound may be weakened slightly, by replacing $N(u)=\lfloor
au-1\rfloor$ with the larger quantity $au-1$, to give
\begin{equation}
\label{Boundmua2}
\rho \ge -\frac{1}{16\pi^3}\int_{\mu}^\infty{\rm d}u\,
\left|\widehat{f^{1/2}}(u)\right|^2
\bigg( u^4+ \frac{2u^3}{a} + \frac{u^2}{a^2}\bigg)\,.
\end{equation}
It clearly differs from the massless Minkowski bound by ${\rm O}(1/a)$ terms. 
The bounds in (\ref{Boundmua1}) and (\ref{Boundmua2}) are plotted in 
Fig.~\ref{fig3} against mass. The difference between these two graphs can 
be further minimised using the approximations below, but at the expense of 
having a more complicated expression for the bound. The corresponding bound 
derived in \cite{pfenning,PFa} is also plotted on the same graph. 

\begin{figure}[t]
\begin{center}
\epsfxsize=6.in
\epsffile{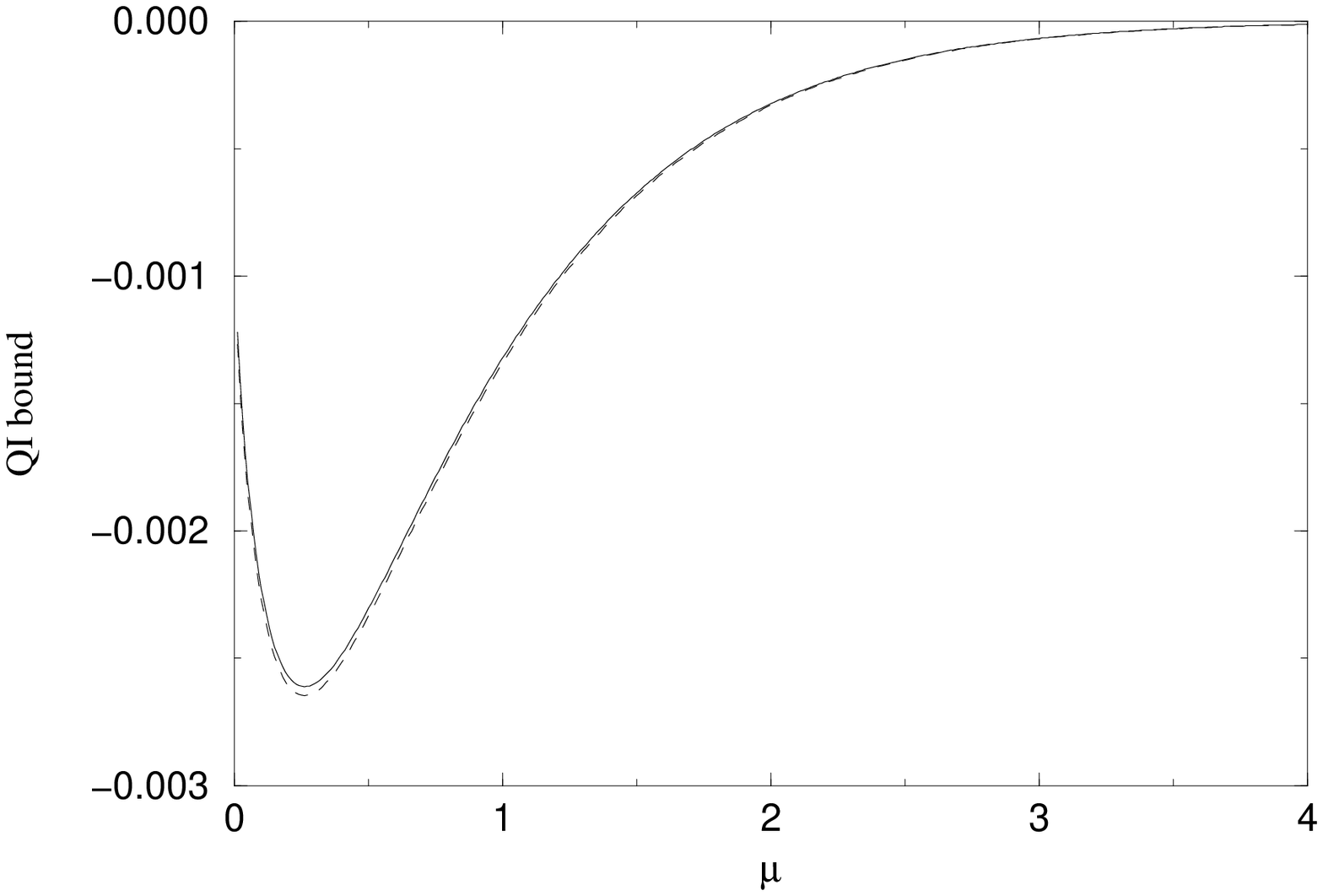}
\caption{Graphs of the QI bounds for the 4D closed universe: 
(\ref{Boundgen1}) using a solid line, and its approximation
(\ref{Boundgen2}) using a dashed line.}
\label{fig4}
\end{center}
\end{figure}

\begin{figure}[t]
\begin{center}
\epsfxsize=6.in
\epsffile{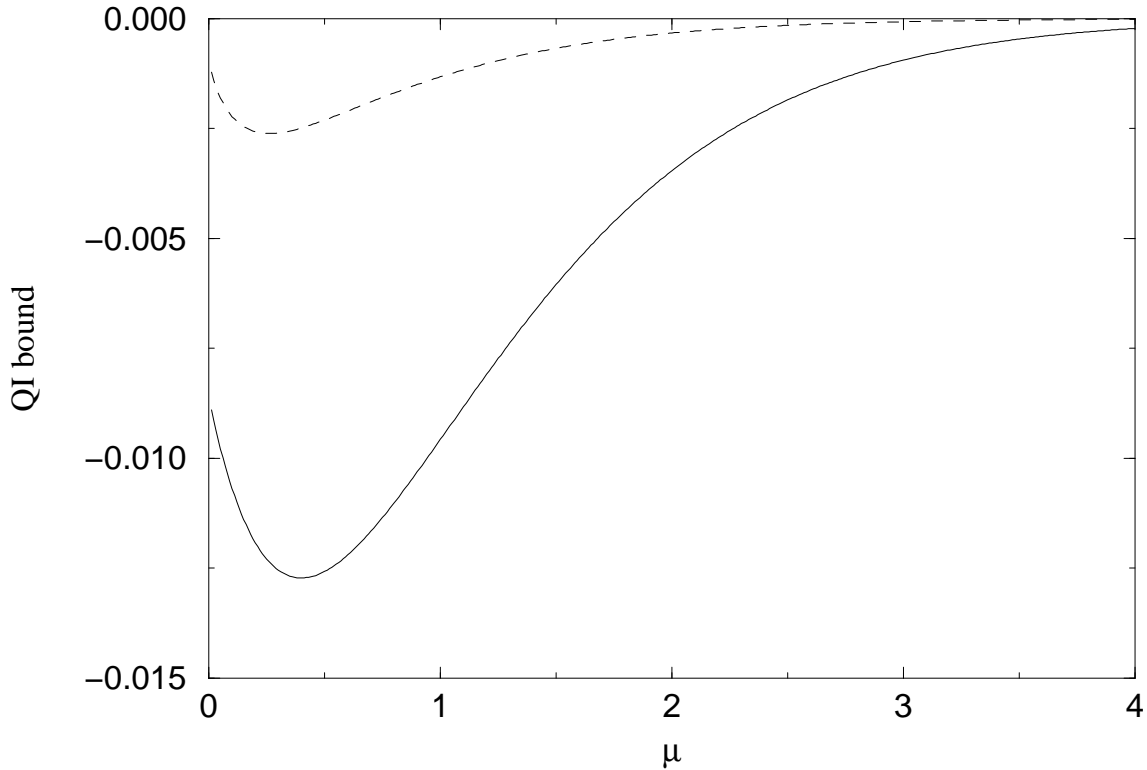}
\caption{Graphs of the QI bounds for the 4D closed universe: 
(\ref{Boundgen1}) using a dashed line, and that of Pfenning and Ford 
[solid line].}
\label{fig5}
\end{center}
\end{figure}

Returning to the general case, we note that (\ref{RWbound}) can be
written as
\begin{equation}
\label{Boundgen1}
\rho\geq-{1\over4\pi^3a^4}\int_\mu^\infty{\rm d}u\,
\left|\widehat{f^{1/2}}(u)\right|^2
\sum_{n=1}^{N'}\,n^2\sqrt{n^2+(a\mu)^2-1}\,,
\end{equation}
where $N'\equiv\lfloor a\sqrt{u^2-\mu^2}\rfloor$. The finite sum can
again be approximated analytically using the trapezoidal rule, now in the form:
\begin{equation}
\sum_{n=1}^Ng(n)=\int_1^N{\rm d}x\,g(x)+{1\over2}\left[g(1)+g(N)\right]
+{N-1\over12}g''(\zeta)\,,
\end{equation}
for some $\zeta\in(1,N)$. From the fact that the second derivative
of $g(x)=x^2\sqrt{x^2+(a\mu)^2-1}$ is non-decreasing in this interval, 
we obtain the inequality
\begin{equation}
\label{Boundgen2}
\rho\geq-{1\over4\pi^3a^4}\int_\mu^\infty{\rm d}u\,
\left|\widehat{f^{1/2}}(u)\right|^2
\left\{\int_1^{N'}{\rm d}x\,g(x)+{1\over2}\left[g(1)+g(N')\right]
+{N'-1\over12}g''(N')
\right\}\,,
\end{equation}
with
\begin{eqnarray}
\int{\rm d}x\,g(x)&=&{1\over4}x\big[x^2+(a\mu)^2-1\big]^{3/2}-{1\over8}
\big[(a\mu)^2-1\big]x\sqrt{x^2+(a\mu)^2-1}\cr
&&\qquad-{1\over8}\big[(a\mu)^2-1\big]^2\ln\Big(x+\sqrt{x^2+(a\mu)^2-1}\Big)\,,
\cr
g''(x)&=&2\sqrt{x^2+(a\mu)^2-1}+{5x^2\over\sqrt{x^2+(a\mu)^2-1}}
-{x^4\over\big[x^2+(a\mu)^2-1\big]^{3/2}}\,.
\end{eqnarray}

The bound in (\ref{Boundgen1}) and its approximation in (\ref{Boundgen2})
are plotted in Fig.~\ref{fig4} against $\mu$, for $a=1$. As can be seen,
the approximation is only very slightly weaker than the exact bound. 
Also plotted in Fig.~\ref{fig5} is the bound obtained in \cite{pfenning,PFa}, 
for comparison with (\ref{Boundgen1}).

\sect{de Sitter space-time}

A convenient static parametrisation of the de Sitter universe is
\begin{equation}
{\rm d}s^2 = -\bigg(1-{r^2\over\alpha^2}\bigg)\,{\rm d}t^2
+\bigg(1-{r^2\over\alpha^2}\bigg)^{-1}{\rm d}r^2+r^2({\rm d}\theta^2
+\sin^2\theta\,{\rm d}\varphi^2)\,,
\end{equation}
with $0\le r\le \alpha$. The surface $r=\alpha$ is the particle horizon 
for an observer located at the origin. In this representation, the mode 
functions for a scalar field with mass $\mu$ and energy $\omega$ are
\begin{equation}
U_{klm}({\bf x})={1\over(4\pi\alpha^2k)^{1/2}}
f_{kl}(z)Y_{lm}(\theta,\varphi)\,,
\end{equation}
where we denote $z\equiv r/\alpha$ and $k\equiv\alpha\omega$. The latter 
continuously parametrises the mode function from zero to infinity, while 
$l$ and $m$ are as in Sec.~\ref{IIIDCU}. The radial function can then be 
solved in terms of the hypergeometric function $F(a,b;c;z)$ as
\cite{higuchi}
\begin{equation}
f_{kl}(z)={\Gamma(b^+_l)\Gamma(b^-_l)\over\Gamma(l+{3\over 2})
\Gamma(ik)}z^l(1-{z^2})^{ik/2}F\bigg(b^+_l,b^-_l;l+{3\over 2};z^2\bigg)\,,
\end{equation}
with
\begin{equation}
b^\pm_l\equiv{1\over 2}\bigg(l+{3\over2}+ik
\pm\sqrt{{9\over4}-\alpha^2\mu^2}\bigg)\,.
\end{equation}

Using the sum rule (\ref{sumrule}) in the quantum inequality (\ref{QIb}), 
we have for an observer at the origin,
\begin{eqnarray}
\rho&\geq&-{1\over64\pi^3\alpha^4}\int_0^\infty{\rm d}\omega
\int_0^\infty{\rm d}k\sum_{l=0}^\infty\,{2l+1\over k}\bigg|{\Gamma(b^+_l)
\Gamma(b^-_l)\over\Gamma(l+{3\over 2})\Gamma(ik)}\bigg|^2\cr&&\quad
\lim_{z\rightarrow0}
\bigg\{{4k^2\over1-z^2}+{1\over z^2}\partial_z\big[z^2(1-z^2)\partial_z\big]
\bigg\}z^{2l}\bigg|F\bigg(b^+_l,b^-_l;l+{3\over 2};z^2\bigg)\bigg|^2
\left|\widehat{f^{1/2}}(\omega+k/\alpha)\right|^2.~~~~~~~
\end{eqnarray}
In fact, only the $l=0$ and $l=1$ terms contribute (cf. Eqs.~(4.126) 
and~(4.127) of~\cite{pfenning}), and the expression may
be simplified to give
\begin{eqnarray}
\label{deSitter}
\rho&\geq&-{1\over8\pi^5\alpha^4}\int_0^\infty{\rm d}\omega
\int_0^\infty{\rm d}k\,\sinh(\pi k)\Big\{(k^2+\alpha^2\mu^2)\big|
\Gamma(b^+_0)\Gamma(b^-_0)\big|^2+4\big|\Gamma(b^+_1)\Gamma(b^-_1)\big|^2
\Big\}\cr&&\hskip1.65in
\times\left|\widehat{f^{1/2}}(\omega+k/\alpha)\right|^2.
\end{eqnarray}

\begin{figure}[t]
\begin{center}
\epsfxsize=6.in
\epsffile{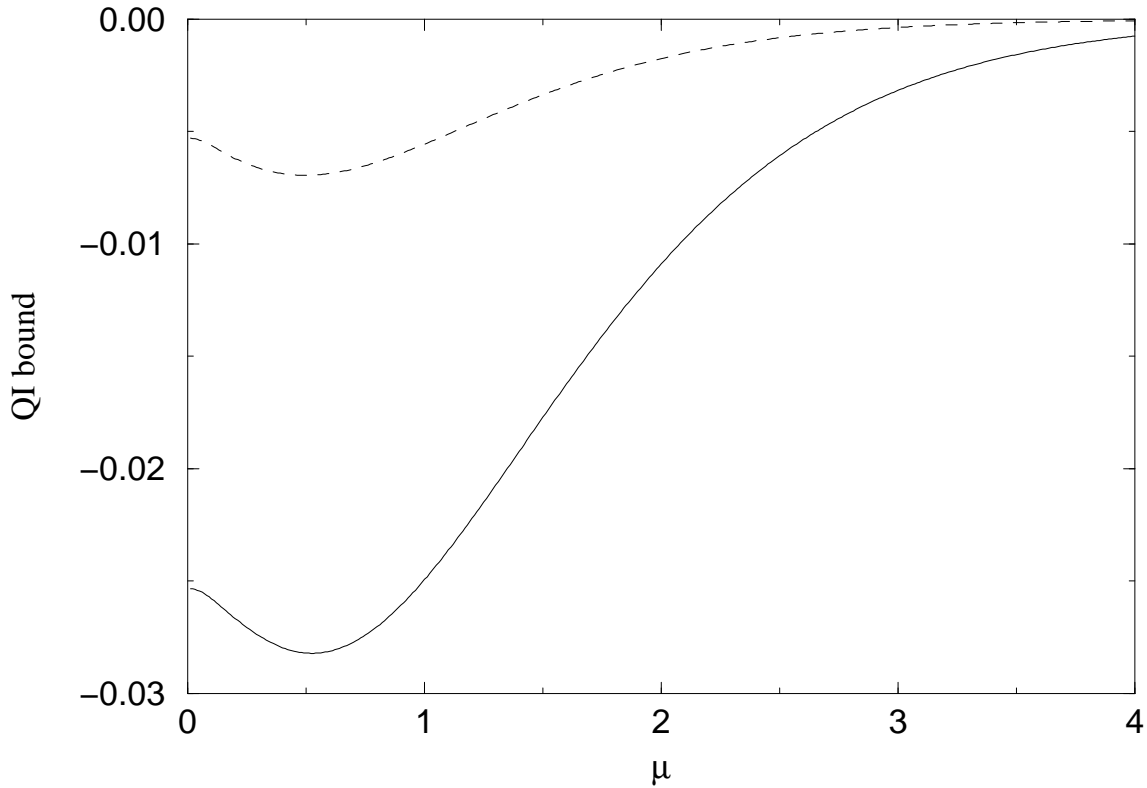}
\caption{Graphs of the QI bounds for de Sitter
space-time: (\ref{deSitter}) using a dashed line, and that of Pfenning 
and Ford [solid line].}
\label{fig6}
\end{center}
\end{figure}

As was noted in \cite{PFb,pfenning}, there are two cases for which the gamma 
functions in (\ref{deSitter}) can be evaluated analytically, namely when
$\mu=0$ and $\sqrt{2}/\alpha$. Assuming the Lorentzian sampling function
(\ref{Lorentzian}) and using (\ref{FT}), we obtain, for the massless case,
\begin{equation}
\rho\geq-{t_0\over2\pi^4\alpha^2}\int_0^\infty{\rm d}\omega
\int_0^\infty{\rm d}\omega^\prime\,(5\omega^\prime+2\alpha^2\omega^\prime{}^3)
K_0\big(t_0(\omega+\omega^\prime)\big)^2.
\end{equation}
Defining the new variables $u=\omega+\omega^\prime$ and $v=\omega^\prime$,
the bound becomes
\begin{equation}
-{t_0\over2\pi^4\alpha^2}\int_0^\infty{\rm d}u\,K_0(t_0u)^2
\int_0^u{\rm d}v\,(5v+2\alpha^2v^3)\,.
\end{equation}
This can be explicitly evaluated using the integral
\begin{equation}
\label{integral}
\int_0^\infty{\rm d}u\,u^{\alpha-1}K_0(t_0u)^2={2^{\alpha-3}
\over t_0^\alpha\Gamma(\alpha)}\Gamma\Big({\alpha\over2}\Big)^4,
\end{equation}
to obtain
\begin{equation}
\label{mydS}
\rho\geq-{3\over32\pi^2t_0^4}{9\over64}\bigg[1+{16\over9}{5\over3}
\Big({t_0\over\alpha}\Big)^2\bigg]\,.
\end{equation}
This bound is at least four times stronger than that obtained in 
\cite{PFb,pfenning}:
\begin{equation}
\label{forddS}
\rho\geq-{3\over32\pi^2t_0^4}\bigg[1+{5\over3}
\Big({t_0\over\alpha}\Big)^2\bigg]\,.
\end{equation}
In the limit $\alpha\rightarrow\infty$ or $t_0\rightarrow0$, we expect
to recover the results for Minkowski space. Indeed, the bound in (\ref{mydS}) 
is then $9\over64$ that in (\ref{forddS}), as was observed in \cite{fewster}.

When $\mu=\sqrt{2}/\alpha$, we similarly obtain the quantum inequality
\begin{equation}
\label{mydS2}
\rho\geq-{3\over32\pi^2t_0^4}{9\over64}\bigg[1+{16\over9}
\Big({t_0\over\alpha}\Big)^2\bigg]\,,
\end{equation}
in contrast to that derived in \cite{PFb,pfenning}:
\begin{equation}
\rho\geq-{3\over32\pi^2t_0^4}\bigg[1+
\Big({t_0\over\alpha}\Big)^2\bigg]\,.
\end{equation}
The bound in (\ref{deSitter}) and that derived in \cite{PFb,pfenning} are 
plotted for general $\mu$, and $\alpha=1$, in Fig.~\ref{fig6}.

We have, in fact, proved that for general $\mu$, the de Sitter bound
(\ref{deSitter}) differs from the Minkowski space bound~(\ref{Mbound2})
by terms no greater than order $\alpha^{-1/2}$ as $\alpha\to\infty$, and
so our results for these cases agree in this limit. 
This estimate involves bounds on the integrand in Eq.~(\ref{deSitter})
which are uniform in $k$ and $\omega$. The proof, which we omit, is
accordingly somewhat technical. It is unclear whether the argument can
be strengthened to show that the deviation is in fact 
${\rm O}(\alpha^{-2})$ in general, as it is for the specific
cases considered in~(\ref{mydS}) and~(\ref{mydS2}).

\sect{Schwarzschild space-time}

As the final example, we shall examine the quantum inequalities in a
black hole space-time. The line element for the Schwarzschild black hole 
of mass $M$ is
\begin{equation}
{\rm d}s^2=-\left(1-{2M\over r}\right){\rm d}t^2
+\left(1-{2M\over r}\right)^{-1}{\rm d}r^2
+r^2({\rm d}\theta^2 + \sin^2\theta\,{\rm d}\varphi^2)\,.
\end{equation}
For simplicity, we shall only consider a massless scalar field in
this space-time. The mode 
functions, in the region exterior to the horizon $r>2M$, take the form
\cite{deWitt}
\begin{eqnarray}
\stackrel{\rightarrow}{U}_{\omega lm}({\bf x})&=&
{1\over(4\pi\omega)^{1/2}}\stackrel{\rightarrow}{R}_l(\omega|r)
Y_{lm}(\theta,\varphi)\,,\cr
\stackrel{\leftarrow}{U}_{\omega l m}({\bf x})&=&{1\over(4\pi\omega)^{1/2}}
\stackrel{\leftarrow}{R}_l(\omega|r)Y_{lm}(\theta,\varphi)\,,
\end{eqnarray}
where, as usual, $\omega$ is the energy of the field and 
$Y_{lm}(\theta,\varphi)$ are the spherical harmonics. 
$\stackrel{\rightarrow}{R}_l(\omega|r)$ and 
$\stackrel{\leftarrow}{R}_l(\omega|r)$
are the outgoing and ingoing solutions to the radial part of the wave 
equation, respectively. Although this equation cannot be solved 
analytically, the asymptotic forms of the solutions are known near the horizon
and at infinity.

Again, using the sum rule (\ref{sumrule}), we see that the quantum inequality
(\ref{QIb}) becomes
\begin{eqnarray}
\label{Schwarz}
\rho&\geq&-{1\over16\pi^3}\int_0^\infty{\rm d}\omega
\int_0^\infty{{\rm d}\omega^\prime\over\omega^\prime}\sum_{l=0}^\infty\,
(2l+1)\,
\bigg\{{\omega^\prime{}^2\over1-{2M\over r}}+{1\over4r^2}\partial_r\Big[
r^2(1-2M/r)\partial_r\Big]\bigg\}\cr
&&\hskip1.65in\times\,\bigg[\Big|\stackrel{\rightarrow}{R}_l(\omega^\prime|r)\Big|^2+
\Big|\stackrel{\leftarrow}{R}_l(\omega^\prime|r)\Big|^2\bigg]
\left|\widehat{f^{1/2}}(\omega
+\omega^\prime)\right|^2.
\end{eqnarray}
In writing this, we are assuming that the mode functions are defined to
have positive frequency with respect to the time-like Killing vector 
$\partial_t$. This is the Boulware vacuum.
Now, in the two regions where $\stackrel{\rightarrow}{R}_l(\omega|r)$ 
and $\stackrel{\leftarrow}{R}_l(\omega|r)$ are known
explicitly, we have \cite{candelas}
\begin{equation}
\sum_{l=0}^\infty\,(2l+1)\Big|\stackrel{\rightarrow}{R}_l(\omega|r)\Big|^2
\simeq\cases{4\omega^2(1-2M/r)^{-1},&$r\rightarrow2M$,\cr
\displaystyle{{1\over r^2}\sum_{l=0}^\infty}(2l+1)\big|B_l(\omega)\big|^2,
&$r\rightarrow\infty$,}
\end{equation}
and
\begin{equation}
\sum_{l=0}^\infty\,(2l+1)\Big|\stackrel{\leftarrow}{R}_l(\omega|r)\Big|^2
\simeq\cases{
\displaystyle{{1\over4M^2}
\sum_{l=0}^\infty}(2l+1)\big|B_l(\omega)\big|^2,&$r\rightarrow2M$,\cr
4\omega^2,&$r\rightarrow\infty$.}
\end{equation}
If we further assume the low-energy condition $2M\omega\ll1$, then 
\cite{jensen}
\begin{equation}
\label{B_l}
B_l(\omega)\simeq{(l!)^3\over(2l+1)!(2l)!}(-4iM\omega)^{l+1}.
\end{equation}
These results can be substituted into (\ref{Schwarz}), and the bound 
explicitly evaluated using the integral (\ref{integral}). However, the 
maximum value of $l$ for which the expansion in (\ref{Schwarz}) is valid 
depends on the order of the leading terms which have been dropped in 
$B_l(\omega)$. If (\ref{B_l}) is exact to ${\rm O}\left[(M\omega)^{l+2}
\right]$, then only the $l=0$ terms should be retained, as the $B_1$ 
contribution would be smaller than the corrections to $B_0$
\cite{PFb,pfenning}.

Near the horizon, the quantum inequality can be expressed in terms 
of the observer's proper time:
\begin{equation}
\tau_0=\bigg(1-{2M\over r}\bigg)^{1\over2}t_0\,,
\end{equation}
as
\begin{equation}
\label{Sbounda}
\rho\geq-{3\over32\pi^2\tau_0^4}\bigg\{  
{1\over24}\bigg({2M\tau_0\over r^2}\bigg)^2\bigg(1-{2M\over r}\bigg)^{-1}
+{9\over64}\bigg[1+\bigg(1-{2M\over r}\bigg)\bigg]+\cdots\bigg\}\,,
\end{equation}
where the ellipsis denotes higher-order terms that have been dropped.
This is to be compared with the result derived in \cite{PFb,pfenning}:
\begin{equation}
\label{Sboundb}
\rho\geq-{3\over32\pi^2\tau_0^4}\bigg\{  
{1\over6}\bigg({2M\tau_0\over r^2}\bigg)^2\bigg(1-{2M\over r}\bigg)^{-1}
+1+\bigg(1-{2M\over r}\bigg)+\cdots\bigg\}\,.
\end{equation}
The bound in (\ref{Sbounda}) is between $9\over64$ and $1\over4$ that in 
(\ref{Sboundb}), at least in the present approximation. Note that in 
either case, the bound becomes arbitrarily negative near the horizon
of the black hole.

On the other hand, the quantum inequality for an observer near infinity becomes
\begin{equation}
\rho\geq-{3\over32\pi^2\tau_0^4}{9\over64}\bigg\{1-{2M\over r}
+\bigg({2M\over r}\bigg)^2\bigg[1+{16\over9}{1\over3}\bigg(
{\tau_0\over r}\bigg)^2\bigg]-\bigg({2M\over r}\bigg)^3
\bigg[1+{16\over9}\bigg({\tau_0\over r}\bigg)^2\bigg]+\cdots\bigg\}\,,
\end{equation}
while the corresponding inequality obtained in \cite{PFb,pfenning} is
\begin{equation}
\rho\geq-{3\over32\pi^2\tau_0^4}\bigg\{1-{2M\over r}+\bigg({2M\over r}\bigg)^2
\bigg[1+{1\over3}\bigg({\tau_0\over r}\bigg)^2\bigg]-\bigg({2M\over r}\bigg)^3
\bigg[1+\bigg({\tau_0\over r}\bigg)^2\bigg]+\cdots\bigg\}\,.
\end{equation}
Again, the former bound is between $9\over64$ and $1\over4$ the latter.
It gives the correct Minkowski space result in the limit $r\rightarrow\infty$
or $M\rightarrow0$.

\sect{Concluding remarks}

In summary, we have derived new quantum inequalities (\ref{QIb}) or 
(\ref{QIc}) on the normal-ordered averaged energy density in static
space-times, that are valid for quite general 
sampling functions. They were then applied to several standard examples 
using the Lorentzian sampling function. (Of course, other 
space-times could readily be considered, such as Rindler 
space, flat space with perfectly reflecting mirrors, and other black holes 
\cite{PFb,pfenning}.) The resulting bounds are stronger than previous results,
and would lead to even tighter constraints on the 
various exotic space-times mentioned at the beginning of the paper.
Before we conclude, a few comments are in order. 

An important question is whether our quantum inequalities are optimal.
This could, for example, be proved by finding a quantum state that
saturates the bound, which would necessarily belong to the kernel of all 
the operators ${\cal O}^\pm(\omega)$ in (\ref{O}). However, it is known 
that our bound, when applied to a massless scalar field in two-dimensional 
Minkowski space, is $1{1\over2}$ times weaker than the optimal value 
obtained by Flanagan \cite{flanagan}. Unfortunately, his 
derivation relies on some special features of two-dimensional massless 
field theory, and does not appear to generalise to other more realistic cases.

An interesting application of our quantum inequality would be to 
the static Morris--Thorne-type wormholes \cite{MT}. Ford and Roman have 
applied the flat-space version of their quantum inequalities to this case, 
and have found that they constrain the size of such wormholes \cite{FR-worm}. 
They justified this procedure by making the sampling timescale much shorter 
than the minimum characteristic curvature scale, so that space-time appears 
locally flat. However, it would be desirable to verify this calculation 
using the full curved space results; this should not be too difficult 
once the form of the scalar field mode functions in the wormhole space-time
have been determined.

\section*{Acknowledgment}

CJF thanks Simon Eveson for useful discussions concerning the use of the
trapezoidal rule in Secs.~\ref{IIIDCU} and~\ref{IVDRWU}.

\appendix
\section{Appendix}

The inequality needed to prove the results in Sec.~\ref{QI} is a 
generalisation of one which was first derived in \cite{fewster} 
using the convolution theorem. Suppose $f$ is a smooth, non-negative 
function of $t$, decaying at least as fast as ${\rm O}(t^{-2})$ for
$t\to \pm\infty$. Let operators $S^\pm$
be defined by 
\begin{equation}
S^\pm={\rm Herm}\,\sum_{\lambda,\lambda^\prime}\Big\{
\widehat{f}(\omega_{\lambda'}-\omega_\lambda)
\overline{q_\lambda} q_{\lambda^\prime}
a_\lambda^\dagger a_{\lambda^\prime}
\pm \widehat{f}(\omega_\lambda+\omega_{\lambda^\prime})
q_{\lambda^\prime} q_\lambda
a_{\lambda^\prime}a_\lambda\Big\}\,,
\label{Spm1}
\end{equation}
where the $q_\lambda$ are complex coefficients,\footnote{For clarity,
we shall denote complex conjugation by an overline in this Appendix.}
and ${\rm Herm}\,X\equiv\frac{1}{2}(X+X^\dagger)$ is the
Hermitian part of an operator $X$. 
We will show that the expectation values $\langle S^\pm\rangle$ obey
\begin{equation}
\label{inequality}
\langle S^\pm\rangle\geq
-\frac{1}{2\pi}\int_0^\infty{\rm d}\omega\sum_\lambda
\left|\widehat{f^{1/2}}(\omega+\omega_\lambda)\right|^2|q_\lambda|^2,
\end{equation}
in any normalised quantum state,
where $f^{1/2}(t)\equiv\sqrt{f(t)}$ is the pointwise square-root 
of $f(t)$. 

To obtain this result, first define a function $g$ by
\begin{equation} 
g(\omega)={1\over\sqrt{2\pi}}\widehat{f^{1/2}}(\omega)\,.
\end{equation}
We have $\overline{g(\omega)}=g(-\omega)$ because $f^{1/2}$ is real, and 
\begin{equation}
(g\star g)(\omega)=\widehat{f}(\omega)\,,
\end{equation}
by the convolution theorem, where $\star$ is given by
\begin{equation}
(g_1\star g_2)(\omega)=\int_{-\infty}^\infty
{\rm d}\omega^\prime\, g_1(\omega-\omega^\prime)g_2(\omega^\prime)\,.
\end{equation}
Next, define the following operators on the space of quantum states:
\begin{equation}
\label{O}
{\cal O}^\pm(\omega)=\sum_\lambda\Big\{
\overline{g(\omega-\omega_\lambda)}q_\lambda a_\lambda\pm
\overline{g(\omega+\omega_\lambda)}\overline{q_\lambda} a_\lambda^\dagger\Big\}\,.
\end{equation}
Using the canonical commutation relations (\ref{ccr}) and symmetrising 
in $\lambda$, $\lambda^\prime$, we obtain
\begin{eqnarray}
\label{OO}
\int_0^\infty{\rm d}\omega\,
{\cal O}^\pm(\omega)^\dagger{\cal O}^\pm(\omega)
&=&\int_0^\infty{\rm d}\omega\sum_{\lambda,\lambda^\prime}
\Big\{
g(\omega-\omega_\lambda) \overline{g(\omega-\omega_{\lambda^\prime})}
\overline{q_\lambda} q_{\lambda^\prime}a_\lambda^\dagger a_{\lambda^\prime}\cr
&&\hskip0.8in+
g(\omega+\omega_\lambda)\overline{g(\omega+\omega_{\lambda^\prime})}
q_\lambda \overline{q_{\lambda^\prime}} a_\lambda a_{\lambda^\prime}^\dagger\cr
&&\hskip0.8in\pm 
g(\omega-\omega_\lambda)\overline{g(\omega+\omega_{\lambda^\prime})}
\overline{q_\lambda q_{\lambda^\prime}} a_\lambda^\dagger 
a_{\lambda^\prime}^\dagger\cr
&&\hskip0.8in\pm 
g(\omega+\omega_\lambda)\overline{g(\omega-\omega_{\lambda^\prime})}
q_\lambda q_{\lambda^\prime}a_\lambda a_{\lambda^\prime}\Big\}\cr
&=&S^\pm+\II\int_0^\infty{\rm d}\omega\sum_\lambda |g(\omega+\omega_\lambda)|^2
|q_\lambda|^2,
\end{eqnarray}
where $S^\pm$ are given by
\begin{equation}
S^\pm = {\rm Herm}\,\sum_{\lambda,\lambda^\prime}\Big\{
F(\omega_{\lambda},\omega_{\lambda'})
\overline{q_\lambda} q_{\lambda^\prime}
a_\lambda^\dagger a_{\lambda^\prime}
\pm G(\omega_\lambda,\omega_{\lambda^\prime})
q_{\lambda^\prime} q_\lambda
a_{\lambda^\prime}a_\lambda\Big\}\,,
\label{Spm2}
\end{equation}
and $F$, $G$ are
\begin{eqnarray}
F(\omega_\lambda,\omega_{\lambda^\prime})&=&
\int_0^\infty{\rm d}\omega\,\Big\{
g(\omega-\omega_\lambda)\overline{g(\omega-\omega_{\lambda^\prime})}
+\overline{g(\omega+\omega_\lambda)}g(\omega+\omega_{\lambda^\prime})
\Big\}\cr
&=& \int_0^\infty{\rm d}\omega\,\Big\{
g(\omega-\omega_{\lambda}) g(-\omega+\omega_{\lambda'}) 
+  g(-\omega-\omega_{\lambda}) g(\omega+\omega_{\lambda'})\Big\}\cr
&=&
\int_{-\infty}^\infty{\rm d}\omega\,g(\omega_{\lambda'}-\omega)
g(\omega-\omega_{\lambda})\cr
&=&(g\star g)(\omega_{\lambda'}-\omega_{\lambda})
=\widehat{f}(\omega_{\lambda'}-\omega_{\lambda})\,,
\end{eqnarray}
\begin{eqnarray}
G(\omega_\lambda,\omega_{\lambda^\prime})&=&
\int_0^\infty{\rm d}\omega\,\Big\{
g(\omega+\omega_\lambda)
\overline{g(\omega-\omega_{\lambda^\prime})}+
\overline{g(\omega-\omega_\lambda)}
g(\omega+\omega_{\lambda^\prime})
\Big\}\cr
&=& \int_0^\infty{\rm d}\omega\,\Big\{
g(\omega+\omega_\lambda) g(\omega_{\lambda'}-\omega)
+g(\omega_\lambda-\omega) g(\omega+\omega_{\lambda'})\Big\}\cr
&=&
\int_{-\infty}^\infty{\rm d}\omega\,g(\omega+\omega_\lambda)
g(\omega_{\lambda'}-\omega)\cr
&=&(g\star g)(\omega_\lambda+\omega_{\lambda^\prime})
=\widehat{f}(\omega_\lambda+\omega_{\lambda^\prime})\,.
\end{eqnarray}
The final equalities show that (\ref{Spm2}) agrees with the 
definition~(\ref{Spm1}). 

Since the left-hand side of (\ref{OO}) is manifestly non-negative, we 
conclude that the expectation value of $S^\pm$ in any normalised quantum 
state must satisfy the inequality
\begin{equation}
\langle S^\pm\rangle\geq-\int_0^\infty{\rm d}\omega
\sum_\lambda |g(\omega+\omega_\lambda)|^2|q_\lambda|^2,
\end{equation}
which is the desired result (\ref{inequality}). Note that the inequality proved in 
\cite{fewster} corresponds to the special case where
the $q_\lambda$ are real and $f$ is an even function of $t$.

\end{document}